\documentclass[aps,pra,twocolumn,superscriptaddress,showpacs]{revtex4-1}
\usepackage[USenglish]{babel} 
\usepackage{graphicx}
\usepackage{dsfont}
\usepackage{rotating}
\usepackage{amsfonts}
\usepackage{xcolor}

\begin{document}

\title{Reversal of relaxation due to dephasing environment}
\author{Marcin Mierzejewski} 
\affiliation{
Institute of Physics, University of Silesia, 40-007 Katowice, Poland}
\author{Janez Bon\v{c}a}
\affiliation{Faculty of Mathematics and Physics, University of Ljubljana, 1000 Ljubljana, Slovenia}
\affiliation{J. Stefan Institute, 1000 Ljubljana, Slovenia}
\author{Jerzy Dajka}
\affiliation{
Institute of Physics, University of Silesia, 40-007 Katowice, Poland}
\affiliation{Silesian Center for Education and Interdisciplinary Research, University of Silesia,   
           41-500 Chorz\'{o}w, Poland}
\begin{abstract} 
 We show that any finite quantum system $S$ can be coupled to a dephasing environment in such a way that  the internal mechanism responsible for relaxation of observables acting on $S$ can be effectively canceled. By adjusting this coupling, the difference between the initial and the long--time expectation values of any observable on $S$  can be tuned to arbitrarily small however nonzero value. This statement is exemplified and visualized by numerical studies of relaxation in a generic  one-dimensional system of interacting fermions.
\end{abstract}

\maketitle


\section{ Introduction}
There is ever-growing research  in the field of dynamics of many-body quantum systems~\cite{polkrev}. One of the central and still unsolved problems is related to equilibration of closed systems, with a particular  emphasis on the relaxation of observables relevant for experimental verification of various theoretical models. Moreover,  in large--enough generic quantum systems one expects that equilibration is equivalent to thermalization. 
 It is commonly accepted that the long--time averages of local observables
in generic systems coincide with expectation values for the statistical Gibbs ensemble \cite{Goldstein2006,sred,Linden2009,Gogolin2011,Riera2012,polkrev}.
 However, our understanding of this process and the condition for its occurrence are far from being complete. For example,  it is known that there is a relation between equilibration and integrability of the system~\cite{Manmana2007,Santos2011,our2013,jpb_int}.  
Generally, nonintegrable  systems are expected to thermalize, while
integrable systems do not approach the Gibbs state but rather  the generalized Gibbs ensemble (GGE) ~\cite{gge,volk,Eckstein2012,Cassidy2011,essler2014,our2014}.
However, there are quantum systems which do thermalize despite being integrable provided that they are prepared in certain states~\cite{rigol2012}. On the other hand, there are nonintegrable quantum systems which do not thermalize~\cite{Gogolin2011} and the role of initial entanglement between subsystems seems to be important.   

According to common intuition, relaxation of observables is a hallmark of {\it irreversibility}~\cite{breu}. If one considers an expectation value of an arbitrary observable $O$ and takes advantage of the spectral decomposition of the Hamiltonian into its eigenstates $H|n\rangle=E_n|n\rangle$, one obtains for an initial
state $|\psi\rangle=\sum_m C_m |m \rangle $:
\begin{eqnarray}\label{evo}
\langle O\rangle(t) &=& \sum_{E_n = E_m} C_n^* C_m \langle n|O|m \rangle\nonumber \\ 
&+&  \sum_{E_n\neq E_m} C_n^* C_m e^{i(E_n-E_m)t}\langle n|O|m \rangle. 
\end{eqnarray}
 Invoking arguments adopted in the context of the eigenstate thermalization hypothesis~\cite{sred}, one expects that  after sufficiently long time, due to destructive interference of oscillating terms in Eq.(\ref{evo}), the observable can relax to its steady state value 
%
\begin{eqnarray}\label{evo1}
\lim_{t\rightarrow\infty} \langle O\rangle(t) &=&
\lim_{\tau \rightarrow\infty}  \frac{1}{\tau}
\int_0^{\tau} \mathrm{ d} t \langle O\rangle(t)  \nonumber \\
&=& \sum_{E_n = E_m} C_n^* C_m \langle n|O|m \rangle.
\label{eq2}
\end{eqnarray}

Although finite closed quantum systems are strictly speaking periodic or quasi--periodic, the irreversibility is still manifested as low (or vanishing) probability of a process which reverses the relaxation. Moreover, if one attaches the quantum system to an infinite environment, transforming a closed  system into an {\it open} system, one expects that the relaxation becomes faster and 'more irreversible' due to the  {\em information loss} or dissipation. 
 This seems at least intuitively indisputable. 
 However, in this paper we present a counter--example to this generally accepted mechanism. We demonstrate that there exist environments 
which cancel internal relaxation mechanisms of an arbitrary quantum system.   
 More precisely, we show that any finite quantum system can be linearly coupled to  a continuous set of harmonic oscillators in such a way that evolution of an arbitrary observable $O$  satisfies:
\begin{eqnarray}\label{evo2}
|\lim_{t\rightarrow\infty}\langle O\rangle(t)- \lim_{t\rightarrow 0}\langle O\rangle(t) | < \epsilon,
\end{eqnarray}
for arbitrary $\epsilon  >0$.  In other words, the relaxation cannot be completely eliminated but can be made arbitrarily inefficient. This prediction is in direct contrast to Eq. (\ref{eq2}). 
The main idea behind this result is that a specially tailored environments may cancel the (destructively interfering) oscillations of the off diagonal matrix elements in Eq. (\ref{evo}). 
 Such environments belong to the class of purely dephasing environments ~\cite{breu}. They  preserve the internal conservation laws of the quantum system, hence they are not generic or even typical.   Applicability of the pure dephasing model has been discussed in various areas of quantum and mathematical physics~\cite{mydef,defaz1,defaz2,assfaza}. Surprisingly, there are also real systems which can be effectively described by such a simple, highly symmetric, model~\cite{defaz}. However, let us stress that pure decoherence  remains credible only at the time scales significantly smaller than the time scale relevant for  exchanging energy between system and its environment~\cite{defaz}. Hence,  applicability
 of the pure dephasing under experimentally accessible conditions is usually at least disputable.

The paper is organized as follows: In the next section we formulate and prove (in subsection A) the central result of our work concerning the reversal of relaxation in dephasing environments. Further,  in subsection B of Section II,  we provide a simple example to illustrate reversal of relaxation  in a generic system of interacting fermions. In Section III we present certain generalizations of the main result. Finally, we conclude our work in last section of the paper.


\section{ Model and main result}   
We consider a quantum system, $S$, described by the Hamiltonian
\begin{eqnarray}
H_S&=&\sum_{n} E_n |n\rangle\langle n|,
\end{eqnarray}
with arbitrary (possibly degenerate) energy spectrum.
The system is coupled to an environment, $B$, of 
 noninteracting bosons
\begin{eqnarray}\label{bos}
H_B&=&\int_0^\infty d\omega \; \omega a^\dagger(\omega)a(\omega),
\end{eqnarray}
where the fields $a(\omega)$ 
 satisfy $[a(\omega),a^\dagger(\omega')]=\delta(\omega-\omega')$~\cite{brat}.  
The details of the  purely dephasing $S$--$B$  coupling   are given by $V_S$ and $V_B$ such that the total Hamiltonian reads~\cite{breu,daj1}
\begin{eqnarray}\label{hamtot}
H&=&H_S\otimes \mathcal{I}_B+\mathcal{I}_S\otimes H_B+ V_S\otimes V_B,
\end{eqnarray}
where
\begin{eqnarray}
V_B&=&\int_0^\infty d\omega g(\omega) \left[a^\dagger(\omega)+a(\omega)\right],
\end{eqnarray}
with a  real valued $g(\omega)$ and 
\begin{eqnarray}\label{cupl}
V_S&=&\sum_n \gamma_n |n\rangle\langle n|.
\end{eqnarray}
Note that $H_S$ and $V_S$ commute. This particular property is a hallmark of a pure dephasing~\cite{breu}.
We assume also that initially the system and its environment are prepared in a separable state
\begin{eqnarray}\label{ini}
\rho(0)=\sum_{n,m} p_{n m}|m\rangle\langle n|\otimes 
|\Omega \rangle\langle\Omega|,
\label{inits}
\end{eqnarray}
where $|\Omega \rangle$ is the bosonic vacuum.  Towards the end
of this paper we discuss more general class of the initial states. 

\subsection{ Proposition}
We consider long--time expectation values of local operators
defined for the quantum system, $O=O_S \otimes \mathcal{I}_B$. 
If the system--environment coupling satisfies the proportionality relation
\begin{eqnarray}\label{cond}
\gamma_m^2- \gamma_n^2& \propto E_m-E_n 
\end{eqnarray}
then the difference 
$ |\lim_{t\rightarrow\infty}\langle O\rangle(t) - \langle O\rangle(0)|$ can
be tuned to arbitrary small value by an appropriate choice of the coupling 
function $g(\omega)$. Most importantly, a single tuning of $g(\omega)$ holds for all local operators $O$.     

  For the  pure dephasing one easily  finds the time-dependent expectation value
\begin{eqnarray}
\langle O\rangle(t) &=& \sum_{n,m} p_{nm} \langle  n|O|m \rangle a_{nm}(t),
\label{evol}
\end{eqnarray}
where 
\begin{eqnarray}
a_{nm}(t)&=& \langle \Omega | \exp[it(E_n+\gamma_n V_B+H_B)] \nonumber \\ 
&&\times \exp[-it(E_m+\gamma_m V_B+H_B)]  | \Omega \rangle.
\end{eqnarray}
Such simple result occurs due to the block--diagonal structure of the Hamiltonian, where each block describes a set of shifted harmonic oscillators: $E_m+\gamma_m V_B+H_B$. An explicit form of $a_{nm}(t)$
has been derived/used many times in  different contexts ranging from mathematical physics~\cite{alidef} up to quantum information~\cite{mydef}.  An explicit form of the amplitude   $a_{nm}(t)$
reads~\cite{daj1}
\begin{eqnarray}
a_{nm}(t)&=& e^{-i(E_m-E_n)t+i(\gamma_m^2-\gamma_n^2)E(t)
-\left(\gamma_m-\gamma_n\right)^2\Lambda(t)}
\label{amn}
\end{eqnarray}
where
\begin{eqnarray}
E(t)&=&\int_0^\infty d\omega \frac{g^2(\omega)}{\omega^2}\left(\omega t-\sin(\omega t)\right),  \\
\Lambda(t)&=&\int_0^\infty d\omega \frac{g^2(\omega)}{\omega^2}\left(1-\cos(\omega t)\right).
\end{eqnarray}
Utilizing the square--integrability of $g(\omega)/\omega$  one finds
from the Lebesgue--Riemann lemma~\cite{bochner}  in the long--time
regime ($t \rightarrow \infty$) that
\begin{eqnarray}
E(t) \rightarrow \tilde{E}t\,\,\,\mbox{and}\,\,\,\, 
\Lambda(t) \rightarrow \tilde{\Lambda} =\mathrm{const}
\end{eqnarray}
where
\begin{eqnarray}
\tilde{E}=\int_0^\infty d\omega g^2(\omega)\omega^{-1}\\
\tilde{\Lambda}= \int_0^\infty d\omega g^2(\omega)\omega^{-2}
\end{eqnarray}

 In order to reduce the effects of the internal relaxation one should tune the system--environment coupling in such a way that the oscillatory 
part in Eq. (\ref{amn}) drops out 
\begin{eqnarray}
E_m-E_n-\tilde{E}(\gamma^2_m- \gamma^2_n)&=&0
\end{eqnarray}
while the exponential term $(\gamma_m- \gamma_n)^{2}
\tilde{\Lambda}$ remains small. In order to show that such a particular choice is indeed possible, we introduce a cut-off frequency ~\cite{breu}, $\omega_c$, for the bosons in the environment and redefine
the coupling function
\begin{eqnarray}
  g(\omega)&=&f(\omega/\omega_c).
\end{eqnarray}
Then, $\tilde{\Lambda} \propto \omega_c^{-1}$ while 
$\tilde{E} \propto  \omega_c^{-0} $. After eliminating the destructive interference of the off--diagonal matrix-elements, the exponential damping may become arbitrarily  small, of the order of $1/\omega_c$ 
\begin{equation}
\log [ a_{mn}(t\rightarrow \infty) ]  \propto - \frac{(\gamma_m-\gamma_n)^2}{\omega_c}.
\end{equation}
While our general scheme does not require any particular
form of $g(\omega)$ the results are most transparent for the standard parametrization of the coupling function~\cite{breu}:
\begin{equation}         
g^2(\omega)=\frac{1}{\Gamma(\mu+1)} \left(\frac{\omega}{\omega_c}\right)^{1+\mu}\exp(-\omega/\omega_c),
\label{g2}
\end{equation}
where $\mu >0$ corresponds to the (mathematically~\cite{alidef}) 
well behaving super--Ohmic environment. In the latter
case $\tilde{E}=1$, while $\tilde{\Lambda}=
(\mu \omega_c)^{-1}$.    

\subsection{ Example}
In order to exemplify our result with the help of a simple but generic case,  we study a one-dimensional  system of $L$ sites
and $L/2$ spin-less fermions with periodic boundary
conditions~\cite{ring1,ring2,ring3,ring4} (i.e. a ring) with a Hamiltonian given by:
\begin{eqnarray}\label{hamfer}
H_S&=&-t_h \sum_{j=1}^L 
\left[ \exp( i \phi) c_{j+1}^\dagger c_j +\mathrm{H.c.} \right] \nonumber \\
&& +  U_{1} \sum_{j=1}^L  n_{j+1}n_j +  U_{2}  \sum_{j=1}^L  n_{j+2}n_j 
\end{eqnarray}
where $n_j=c_j^\dagger c_j$, $t_h$ is the hopping integral, whereas $U_1$ and $U_2$  describe first-- and second--nearest--neighbor interactions, respectively. We take $t_h$  as the energy unit $t_h=1$, whereas time is expressed in units
of $\hbar/t_h$. We take also $U_1=1.4$ and $U_2=1$. For such parameters $H_S$ describes a generic metal characterized by a featureless response to electromagnetic field and a normal diffusive transport.  Moreover, such system thermalizes even when being decoupled from its surrounding \cite{our2013}.  Numerical studies have been
carried out for $L \le 18$. The reason behind introducing $U_2$ is to stay away from 
the integrable case, which shows anomalous relaxation \cite{gge,Eckstein2012,Cassidy2011,my3} and charge transport
\cite{my1,my2,Tomaz2011,Marko2011,Sirker2009,Robin2011}. 

The mechanism of the reversal of relaxation (RR)  holds either for all observables or for none.  In the following, we show numerical results
for a  particle current 
\begin{eqnarray}\label{curfer}
O_S &=&\sum_{j=1}^L i \exp( i \phi) c_{j+1}^\dagger c_j +
\mathrm{H.c.},
\end{eqnarray}
since $\langle O_S \rangle$ vanishes whenever S is in equilibrium, hence large  or even non--zero values of this observable 
imply that the system is in a non--thermal state.
We consider a typical super-Ohmic environment with  $\mu=1$ [see  Eq.(\ref{g2})] when 
\begin{eqnarray}
E(t)  =  t \; \frac{\omega^2_ct^2}{1+\omega^2_ct^2}, \quad
\Lambda(t)  =  \frac{1}{\omega_c} \; \frac{\omega^2_ct^2}{1+\omega^2_ct^2}.
\end{eqnarray}
While our qualitative conclusions do not depend on the initial state, we assume that the system is initially in a pure
state $ p_{nm}=C_n C_m $ [see Eq. (\ref{inits})], where $C_n=\mathrm{const}>0$  if $\sum_m \mathrm{Re} \langle n|O_S|m\rangle  >0$ and $C_n=0$ otherwise. Such a state gives large initial current and has non--zero 
projection on large number (approximately one half) of energy eigenstates. Consequently, it leads to very clear numerical results for the relaxation of the particle current.

\begin{figure}[!b]
\begin{center}
\includegraphics[scale=1.4]{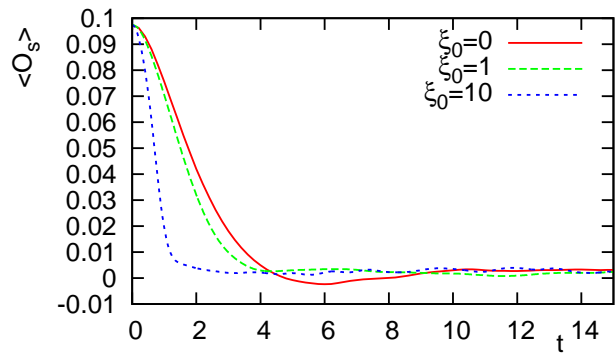}
\end{center}
\caption{(color online). Time dependence of the particle
current in closed generic system ($\xi_0=0$) and for generic
(random) coupling $ \gamma_m^2 =\xi_m$, where $\xi_m$ is  random
variable with flat distribution in $[0,\xi_0]$. Here $\omega_c=1$ has
been assumed.   
}
\label{fig1}
\end{figure}

In Fig. \ref{fig1} we present the relaxation of current flowing in an isolated ring, i.e. in a closed system decoupled 
from any environment. Numerical results show that the fermionic system under consideration is generic and large enough 
so it relaxes (close) to equilibrium due to internal scattering processes even in the absence of any (dephasing) environment.
In the same figure we show relaxation in the presence of a typical dephasing described by the following choice  $\gamma^2_m =\xi_m$ [c.f. Eq.(\ref{cupl})], where $\xi_m$ is a  random
variable with a uniform distribution in an interval $[0,\xi_0]$. This case is very different from the tuned coupling described by  Eq.(\ref{cond}). We notice that such a typical dephasing additionally accelerates the relaxation. Hence the model under consideration reproduces the well expected and intuitive results. 

\begin{figure}[!b]
\begin{center}
\includegraphics[scale=1.4]{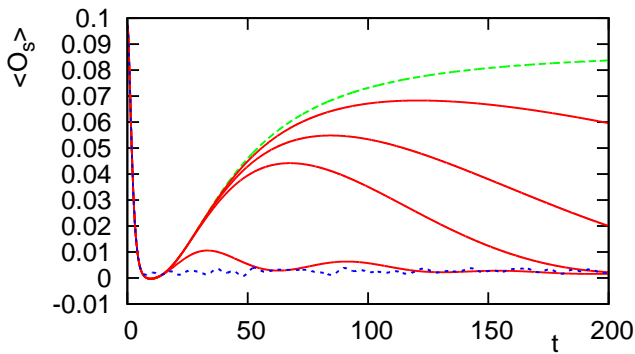}
\end{center}
\caption{(color online). Time dependence of the particle
current for a partially tuned coupling $ \gamma_m^2 =E_m-E_0 + \xi_m$, where $\xi_m$ is  random
variable with flat distribution in $[0,\xi_0]$. $\omega_c=0.1$ has
been assumed. The curves from the top (green) to the bottom (blue) are for $\xi_0=$0, 0.01, 0.02, 0.03, 0.1, and 1, respectively}   
\label{fig2}
\end{figure}

The situation dramatically changes in the  case of  the fine--tuned  coupling between the ring and its bosonic environment satisfying Eq.(\ref{cond}). According to the Proposition we expect reversal of the relaxation, i.e. after sufficiently long evolution time the expectation value of the current should approach its initial value. However, the  realistic systems are never perfect and one can expect that  the condition in Eq.(\ref{cond}) is satisfied  at most approximately.
Let us assume that the the most optimal achievable tuning is given by
\begin{eqnarray}\label{condran}
\gamma_m^2 &=&E_m-E_0+\xi_m
\end{eqnarray}
where $\xi_m\in[0,\xi_0]$ is again a uniformly distributed  random variable. It describes a degree of quenched or frozen disorder present in our system due its imperfect preparation. The time evolution of current flowing in the ring for different values of $\xi_0$ is presented in Fig. \ref{fig2}. In particular, for an ideal case $\xi_m\equiv 0$ condition in Eq.(\ref{cond}) is satisfied and the RR clearly occurs. For small but non-vanishing values of $\xi_0$ there is still a wide time-window with significant degree of RR as indicated in Fig. \ref{fig2}. This nonmonotonic behavior with a wide plateau represents a hallmark of RR that could possibly be observed in simple quantum systems provided their couplings to the environments could be appropriately tuned.

\begin{figure}[!b]
\begin{center}   
\includegraphics[scale=1.4]{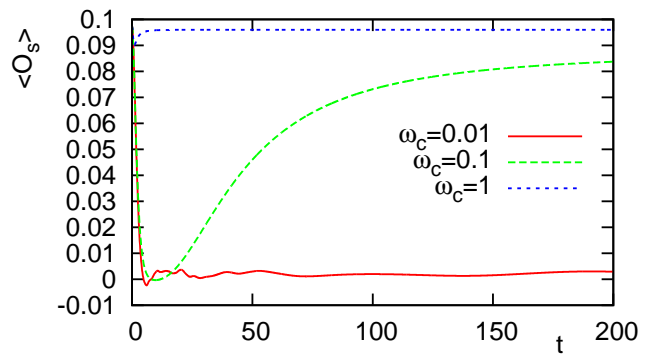}
\end{center}
\caption{(color online). Time dependence of the particle
current for a perfectly tuned coupling $ \gamma_m^2 =E_m-E_0$ and
various $\omega_c$.}   
\label{fig3}
\end{figure}

Finally, we discuss the role of the bosonic characteristic (cut-off)  frequency $\omega_c$. When this quantity
is too small the coupling to the environment is not beneficial any more. Although RR is still possible, the off-diagonal matrix elements $a_{nm}$ decay in time due to the exponential terms in Eq. (\ref{amn}). Results shown in Fig. {\ref{fig3}} for a perfectly tuned coupling $ \gamma_m^2 =E_m-E_0$ indicate that 
a clear RR effect is observed already when $\omega_c$ is one order of magnitude smaller that the typical energy scale of the quantum system, $\sim t_h$. 
If  $\omega_c$ is of the same order of magnitude or even larger then the exponential damping becomes hardly visible. 

\section{Generalizations}
The  assumptions behind the Proposition are not very restrictive, nevertheless they limit applicability of our main result to a certain class of problems. Here we discuss which assumptions can be relaxed  without a significant modification of Eq.(\ref{evo2}). 
First, one may consider a more general class of initial separable states, Eq.(\ref{ini}),  where bosonic subsystem is in the coherent state $|\Xi_\zeta\rangle=D(\zeta)|\Omega\rangle$, where $D$ is the displacement operator~\cite{breu,PhysRevA.79.012104} where $\zeta(\omega)$ is square--integrable and the product $\zeta(\omega)g(\omega)$ satisfies Lebesgue--Riemann lemma. This condition is satisfied, e.g. for  a  non--negative and square--integrable $\zeta(\omega)$.  Of course, for $\zeta(\omega)\equiv 0$ one arrives back to the Proposition. 

It is known that the initial system--environment entanglement can affect thermalization processes~\cite{Gogolin2011}. A simple example of an entangled pure initial (not normalized) state is given by $|n\rangle\otimes|\Xi_\zeta\rangle+|m\rangle\otimes|\Xi_\chi\rangle$ where $|\Xi_\chi\rangle$ and  $|\Xi_\zeta\rangle$ are noncolinear. Such a simple initial preparation  of the pure dephasing models still allows us to find the exact time dependence of the total  system$+$environment composite~\cite{daj_ent}. Again, the reversal of relaxation can occur provided that both $\zeta(\omega)g(\omega)$ and $\chi(\omega)g(\omega)$ satisfy the Lebesgue--Riemann lemma.

\section{ Conclusions} 
The effect of an environment attached to a small system may occasionally  be  counter--intuitive or even unpredictable. Let us mention only two notable examples: stochastic resonance~\cite{sr} (both classical and quantum) and 
environment--induced entanglement~\cite{indent}.      
In our work we present another example of a counter--intuitive effect, which is the reversal of relaxation due to 
a coupling to the environment.
 We show that relaxation of an arbitrary observable acting on a finite quantum system can effectively be reversed by attaching the system to an infinite super-Ohmic dephasing reservoir. 
Since this mechanism requires a fine--tuning of the coupling between the energy levels and the bosonic bath, we expect that it could possibly be realized
in  simple quantum  systems  rather than in complex generic cases. For the latter systems our finding will most probably remain only a matter of principle, in particular since such a coupling represents a non--local interaction. However, even for an imperfect tuning of the interaction between the system and the environment, one still finds a partial reversal of relaxation.
This mechanism may be realized in open quantum systems  when the time scale of energy exchange with the  environment is large when compared to other time scales of the system. Such an approximation has been effectively  applied to describe quantum optical systems coupled to a single electromagnetic mode~\cite{defaz}.   
The  effect studied in our work shows up as a broad time--window in which the expectation values of observables are close to the values in the initial state. Even though the proposed mechanism for the reversal of relaxation is at this stage rather theoretical, it may have an important impact in the area of quantum computing where one of the greatest challenges is controlling or removing quantum decoherence  among interacting quantum systems.

\section*{Acknowledgments}
We  acknowledges stimulating discussions with S. A. Trugman.
This work has been supported by Polish National Science Center under the grant DEC-2013/09/B/ST3/01659 (M.M. and J.D.) and by the P1-0044 project of ARRS (J.B.).  The work by J.B. was in part performed at the Center for Integrated Nanotechnologies, a US Department
of Energy, Office of Basic Energy Sciences user facility.

\bibliography{bibligraphy.bib}

\end{document}